\documentstyle[11pt,newpasp,epsf,twoside]{article}
\def\edcomment#1{\iffalse\marginpar{\raggedright\sl#1\/}\else\relax\fi}
\reversemarginpar

\begin{document}
\title{Dynamical Instabilities in Extrasolar Planetary Systems}
\author{Eric B.\  Ford}
\affil{Department of Astrophysical Sciences, Princeton University, Peyton Hall - Ivy Lane, Princeton, NJ 08544}
\author{Frederic A.\ Rasio and Kenneth Yu}
\affil{Northwestern University, Dept. of Physics \& Astronomy, 2145 Sheridan Rd., Evanston, IL 60208-0834}

\begin{abstract}
Instabilities and strong dynamical interactions between multiple giant
planets have been proposed as a possible explanation for the
surprising orbital properties of extrasolar planetary systems.  In
particular, dynamical instabilities seem to provide a natural
mechanism for producing the highly eccentric orbits seen in many
systems.
Previously, we performed numerical integrations for the dynamical
evolution of planetary systems containing two giant planets of equal
masses initially in nearly circular orbits very close to the dynamical
stability limit.  We found the ratio of collisions to ejections in
these simulations was greater than the ratio of circular orbits to
eccentric orbits among the known extrasolar planets.  Further, the
mean eccentricity of the planets remaining after an ejection was
larger than the mean eccentricity of the known extrasolar planets.
Recently, we have performed additional integrations, generalizing to
consider two planets of unequal masses.  Our new simulations reveal
that the two-planet scattering model can produce a distribution of
eccentricities consistent with the observed eccentricity distribution
for plausible mass distributions.  Additionally, this model
predicts a maximum eccentricity of about $0.8$, in agreement with observations.  
Early results from simulations of three equal-mass planets also reveal a
reduced frequency of collisions and a broad range of final
eccentricities for the retained inner planet.
\end{abstract}

\section{Background}

The known extrasolar planets (Fig.~1 left) can be roughly divided into two
groups: those with short-period, nearly circular orbits ($a\la
0.07\,$AU) and those with wider and more eccentric orbits ($a\ga
0.07\,$AU).
Many of the short-period planets, like their prototype 51~Peg, are so
close to their parent star that tidal dissipation would have likely
circularized their orbits, even if they were originally eccentric
(Rasio {\it et al.\/}~1996).  Thus, their small observed
eccentricities do not provide a good indicator of their dynamical
history.  
However, the large eccentricities of most planets with longer periods 
require an explanation.  A planet that would have formed from a
protoplanetary disk in the standard manner is unlikely to have
developed such a large eccentricity, since dissipation in the disk
tends to circularize orbits. Dynamical instabilities leading to the
ejection of one planet while retaining another planet of comparable
mass could explain the observed distribution of
eccentricities.  

In previous papers (Rasio \& Ford 1996; Ford, Havlickova, \& Rasio
2001), we conducted Monte Carlo integrations of planetary systems
containing two equal mass planets initially placed just inside the
Hill stability limit (Gladman 1993).  We numerically integrated the
orbits of such systems until there was a collision, or one planet was
ejected from the system, or some maximum integration time was reached.  
The two
most common outcomes were collisions between the two planets,
producing a more massive planet in a nearly circular orbit
between the two initial orbits, and ejections of one planet
from the system while the other planet remains in a
tighter orbit with a large eccentricity.  The relative frequency of
these two outcomes depends on the ratio of the planet radius to the
initial semi-major axis.
While this model could naturally explain how planets might acquire
large eccentricities, upon comparing our results to the observed
planets we found two important differences.  First, for the relevant
radii and semi-major axes, we found that collisions are more frequent
in our simulations than nearly circular orbits among the presently
known extrasolar planets.  Second, of the systems which ended with one
planet being ejected from the system, the eccentricity distribution of
the remaining planet was concentrated in a narrow range which is
greater than the typical eccentricity of the known extrasolar planets
(See Fig.~1, right).
Here we present results for new simulations involving two planets with
unequal masses and three planets with equal masses.

\begin{figure}[htb]
\plottwo{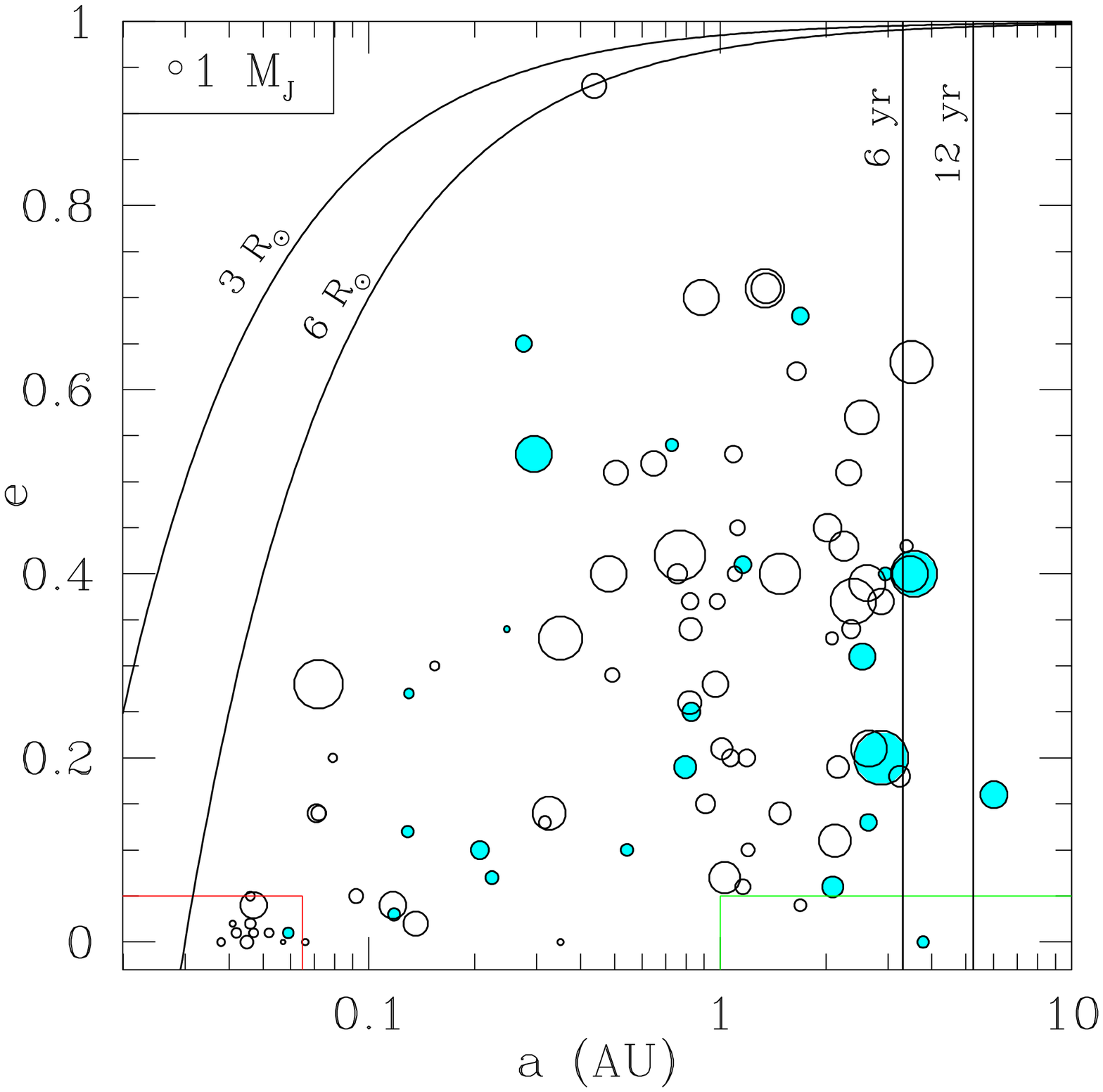}{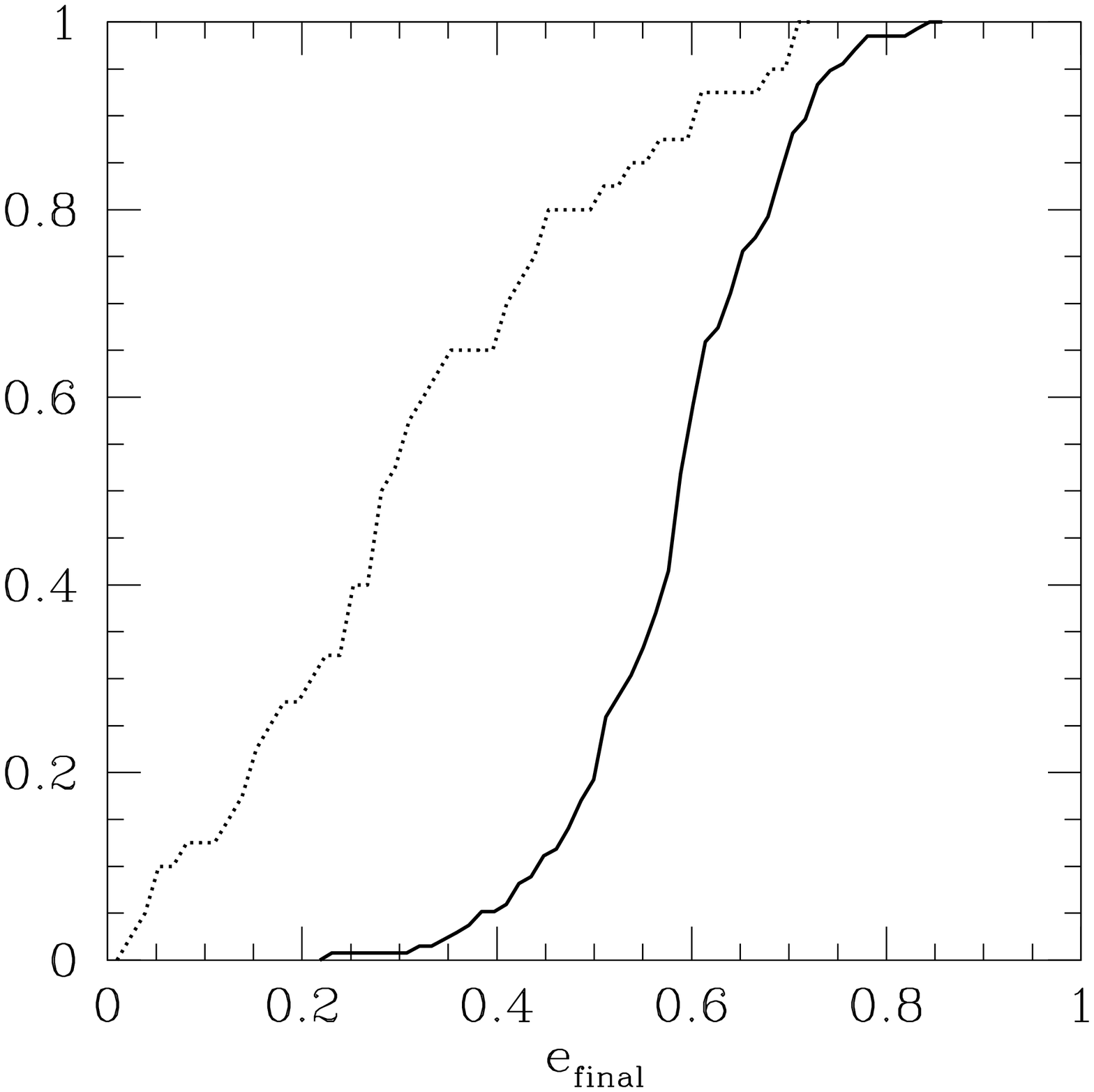}
\caption{Left:  Orbital eccentricity versus semi-major axis of known extrasolar
planets. The eccentricity of planets in very short period orbits is
limited by tidal damping.  Nearly all of the remaining planets have
large eccentricities.  The area of the circle is proportional to $m \sin i$. 
Planets represented by shaded circles are part of a multiple planet system.
Right: Cumulative distribution of eccentricity. The solid line shows
the eccentricity distribution of the remaining planet after the other
planet has been ejected from the system in our previous simulations
with two planets of equal mass.  The dotted line shows the
eccentricity distribution of the known extrasolar planets (excluding
those with $P < 10$d).  Among systems resulting in an ejection, the
final eccentricity is larger than typical for the observed extrasolar
planets.  }
\end{figure}

\section{Two Planets, Unequal Masses}

\subsection{Numerical Setup}

The new simulations use the mixed variable symplectic algorithm (Wisdom
\& Holman 1991) modified to allow for close encounters between planets as
implemented in the publicly available code Mercury (Chambers 1999).
The results presented below are based on $\sim 10^4$ numerical
integrations.  

Throughout the integrations, close encounters between any two bodies
were logged, allowing us to present results for any values of the
planetary radii using a single set of orbital integrations.  We
consider a range of radii to allow for the uncertainty in both the
physical radius and the effective collision radius allowing for
dissipation in the planets.  When two planets collided, mass and
momentum conservation were used to compute the final orbit of the
resulting single planet.

Each run was terminated when one of the following four conditions was
encountered: (i) one of the two planets became unbound (which we
defined as having a radial distance from the star of $1000\,
a_{1,\mathrm{init}}$); (ii) a collision between the two planets
occurred assuming $R_i/a_{1,\mathrm{init}} = R_{\rm
min}/a_{1,\mathrm{init}} = 0.1\, R_{\rm Jup}/5\,{\rm AU}=
0.95\times10^{-5}$; (iii) a close encounter occurred between a planet
and the star (defined by having a planet come within $r_{\rm
min}/a_{1,\mathrm{init}}= 10\,R_\odot/1\,{\rm AU}= 0.06$ of the star);
(iv) the integration time reached $t_{\rm max}= 5\cdot 10^6 - 2 \cdot
10^7$ depending on the masses of the planets.
These four types will be referred to as ``collisions,'' meaning a
collision between the two planets, ``ejections,'' meaning that one
planet was ejected to infinity, ``star grazers,'' meaning that one
planet had a close pericenter passage, and ``two planets,''
meaning that two bound planets remained in a (possibly new)
dynamically stable configuration.

Our numerical integrations were performed for a system containing two
planets, with mass ratios $10^{-4} < m_i/M < 10^{-2}$, where $m_i$
is the mass of one of the planets and $M$ is the mass of the central
star.  A mass ratio of $m/M \simeq 10^{-3}$ corresponds to $m\simeq1\,M_
{\rm Jup}$ for $M=1\,M_\odot$.
The 
% initial semimajor axis of the inner planet ($a_{1,\mathrm{init}}$) was set to unity and the 
initial semimajor axis of the outer planet
($a_{2,\mathrm{init}}$) was drawn from a uniform distribution ranging
from $0.9 \cdot a_{1,\mathrm{init}} \left( 1+\Delta_c \right) $ to 
$a_{1,\mathrm{init}} \left( 1+\Delta_c\right)$, where $1 + \Delta_c$ is
the critical ratio above which Hill stability is guaranteed for
initially circular coplanar orbits (Gladman 1993).
The initial eccentricities were distributed uniformly in the range
from 0 to 0.05, and the initial relative inclination in the range
from $0^{\circ}$ to $2^{\circ}$. All remaining angles (longitudes and
phases) were randomly chosen between 0 and $2\pi$.  Throughout this
paper we quote numerical results in units such that $G=a_{1,\mathrm{init}}=M=1$. In
these units, the initial orbital period of the inner planet is
$P_1\simeq 2\pi $.

%\begin{figure}[htb]
%\plottwo{ExEject.ps}{ExCollide.ps}
%\caption{Examples of two planet scattering: For each planet we show the
%semi-major axis, pericenter, and appocenter distance versus time.  One
%common outcome is one planet being ejected from the system and the
%other planet remaining in an eccentric orbit (left).  Another common
%outcome is the two planets colliding to form a more massive planet in
%a nearly circular orbit (right).  }
%\end{figure}

\subsection{Results}

\subsubsection{Branching Ratios}

For the range of $m_i/M$ considered (up to $10^{-2}$), the frequency of
the various outcomes is relatively insensitive to the ratio of the
total planet mass to the stellar mass.  However the frequencies of the
different outcomes are significantly effected by the planetary mass ratio
and by the planetary radii. Collisions are most
common for large planetary radii and mass ratios near unity, while
ejections are most common for small planetary radii and mass ratios
far from unity.  The frequency of outcomes does not depend on whether
the more massive planet is initially in the inner or outer orbit.  In
Fig.~2 (left), we show the frequency of ejection or collision
as a function of the planetary
radii with solid and dashed lines, respectively.  The black lines
show the result for equal mass planets.  The other colors are for
different mass ratios.  Ejections become more common and collisions
less frequent as the planetary mass ratio departs from unity.

%Two Planet Equal Mass Branching Ratio caption
%Branching ratio versus planet radius: The relative frequency of
%collisions (dotted line) and ejections (dashed line) depends on the
%planet's radius relative to its semi-major axis.  Comparing to the
%semi-major axes and expected radii of the observed extrasolar planets,
%these simulations produce an overabundance of collisions resulting in
%nearly circular orbits.

\subsubsection{Collisions}

Collisions leave a single, larger planet in orbit around the star.
The energy in the center-of-mass frame of the two planets is much
smaller than the binding energy of a giant planet.  Therefore, we
model the collisions as completely inelastic and assume that the two
giant planets simply merge together while conserving total momentum
and mass.  Under this assumption, we have calculated the distributions
of orbital parameters for the collision products.  The final orbit has
a semi-major axis between the two initial semi-major axes, a small
eccentricity, and a small inclination.  While collisions between
planets may affect the masses of extrasolar planets, a single
collision between two massive planets does not cause significant orbital
migration or eccentricity growth if the planets are initially on low-eccentricity, 
low-inclination orbits near the Hill stability limit.

\subsubsection{Ejections}

\begin{figure}[htb]
\plottwo{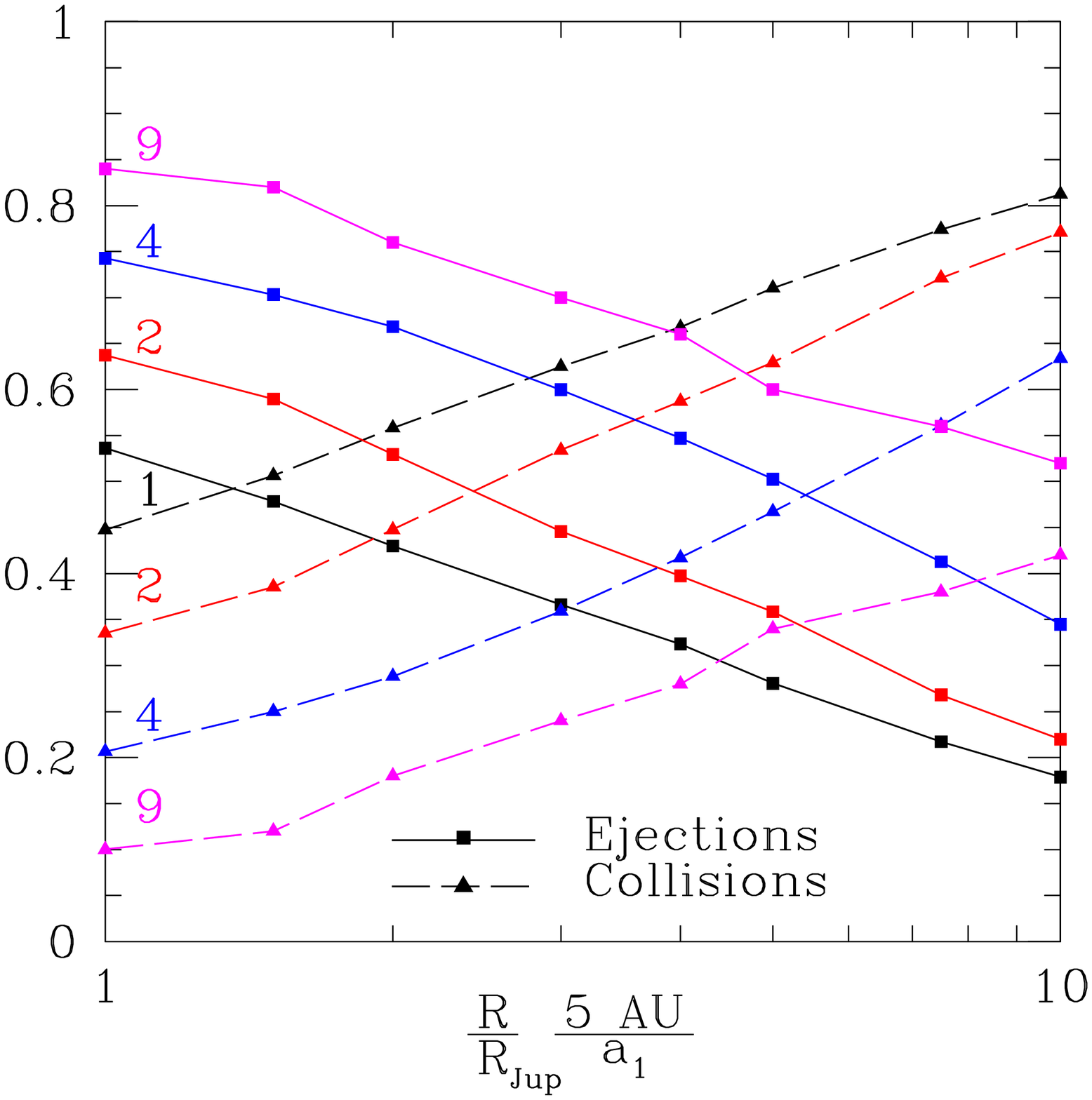}{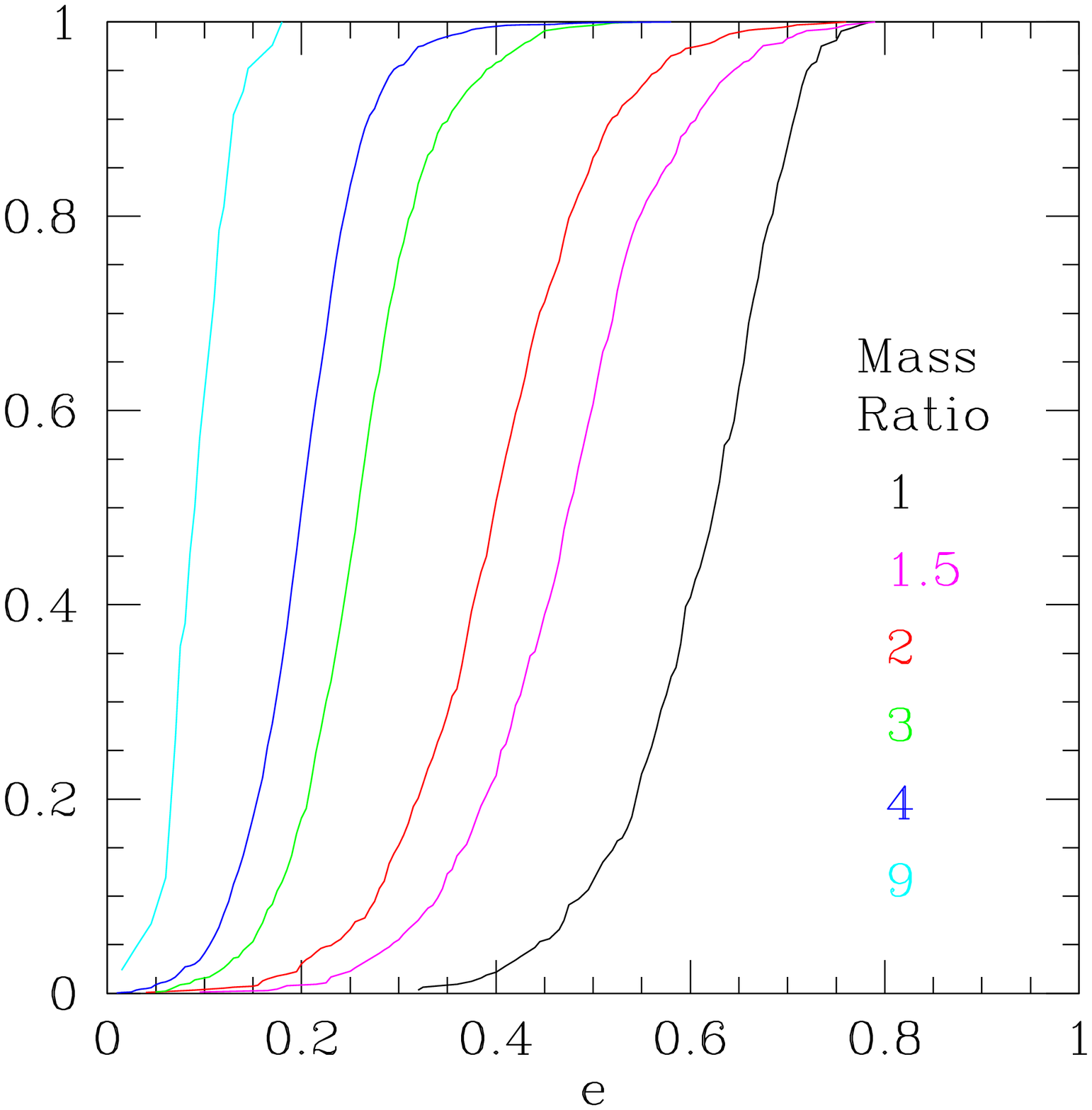}
\caption{Left: Branching ratio versus planet radius. The frequency of collisions
(dashed lines) and ejections (solid lines) depends on the planet's
radius relative to the semi-major axis (x-axis) and the planetary mass
ratio (different colors).  The fraction of integrations
resulting in one planet being ejected and the other planet remaining
in an eccentric orbit is increased when the two planetary masses
differ. 
Right: Cumulative distribution of eccentricity. Each line shows the
cumulative distribution of the eccentricity of the remaining planet
after the other planet has been ejected from the system in our
simulations for a particular mass ratio.  The distributions are not
sensitive to the total mass of the planets or to which planet is more
massive.    }
\end{figure}

When one planet is ejected from the system, the less massive planet is
nearly always ejected if $\frac{m_<}{m_>}\le 0.8$, where $m_<$ and
$m_>$ refer to the masses of the less and more massive planets,
respectively.  Since the escaping planet typically leaves the system
with a very small (positive) energy, energy conservation sets the
final semimajor axis of the remaining planet slightly less than
\begin{equation}
\frac{a_f}{a_{1}} \simeq \frac{a_2 \, m_< }{a_1 \left( m_1+m_2 \right) + \left(a_2-a_1\right) m_1}.
\end{equation}
Thus, the ejection of one of two equal mass planets results in the
most significant reduction in the semi-major axis, but is limited to
$\frac{a_f}{a_{1,\mathrm{init}}} \ge 0.5$.  The remaining planet acquires a
significant eccentricity, but its inclination typically remains
small.  The eccentricity distribution for the remaining planet is
not sensitive to the sum of the planet masses, but depends significantly
on the mass ratio.
In Fig.~2 (right) we show the cumulative distributions
for the eccentricity after a collision for different mass ratios in
different colors.  While any one mass ratio results in a narrow range
of eccentricities, a distribution of mass ratios would result in a
broader distribution of final eccentricities.  Also note that there is
a maximum eccentricity which occurs for equal mass planets.  Thus, the
two planet scattering model predicts a maximum eccentricity of about
0.8 independent of the distribution of planet masses.  While this
compares favorably with the presently known planets, future
observations will certainly test this prediction.

\subsubsection{Stargrazers}

In a small fraction of our numerical integrations ($\sim 3\%$) one
planet underwent a close encounter with the central star.  Due to the
limitations of the numerical integrator used, the accuracy of our
integrations for the subsequent evolution of these systems can not be
guaranteed.  Moreover, these systems could be affected by additional
forces (e.g., tidal forces, interaction with the quadrupole moment of the star)
that are not included in our simulations and would depend
on the initial separation and the radius of the star.  Nevertheless,
our simulations suggest that for giant planets with initial semimajor
axes of a few AU, the extremely close pericenter distances necessary
for tidal circularization around a main sequence star (leading to the
formation of a 51-Peg-type system) are possible, but rare.

\subsection{Discussion}

We now consider whether the the two planet scattering model could
produce a distribution of eccentricities consistent with that of known
extrasolar planets with periods longer than 10 days.  In Fig.~3 (left)
we show the cumulative distribution for the
eccentricity of the remaining planet in our simulations.  Here we have
chosen a simple distribution of planet masses, $P(m) \simeq m^{-1}$
for $1 M_{\rm Jup} \le m \le 10 M_{\rm Jup}$, which is consistent with the
mass distribution of the observed extrasolar planets (Tabachnik \&
Tremaine 2001).  The red dashed line is just for systems which
resulted in an ejection, while the green dotted line includes systems
which resulted in a collision.  Both are reasonably close to the
observed distribution shown by the solid black line.  It should be noted that the
mass distribution used for this model has two cutoffs.  While the $10
M_{\rm Jup}$ cutoff is supported by observations, the lower cutoff is not
constrained by present observations and is unlikely to be physical.
Alternative mass distributions differing for $m \le 1 M_{\rm Jup}$ could
alter this result.  A detailed comparison to observations would require the
careful consideration of observational selection effects, as well as the unknown
initial mass distributions.  Nevertheless, we find that the two-planet
scattering model is able to reproduce the eccentricity distribution of
the known extrasolar planets for plausible mass distributions.

\begin{figure}[htb]
\plottwo{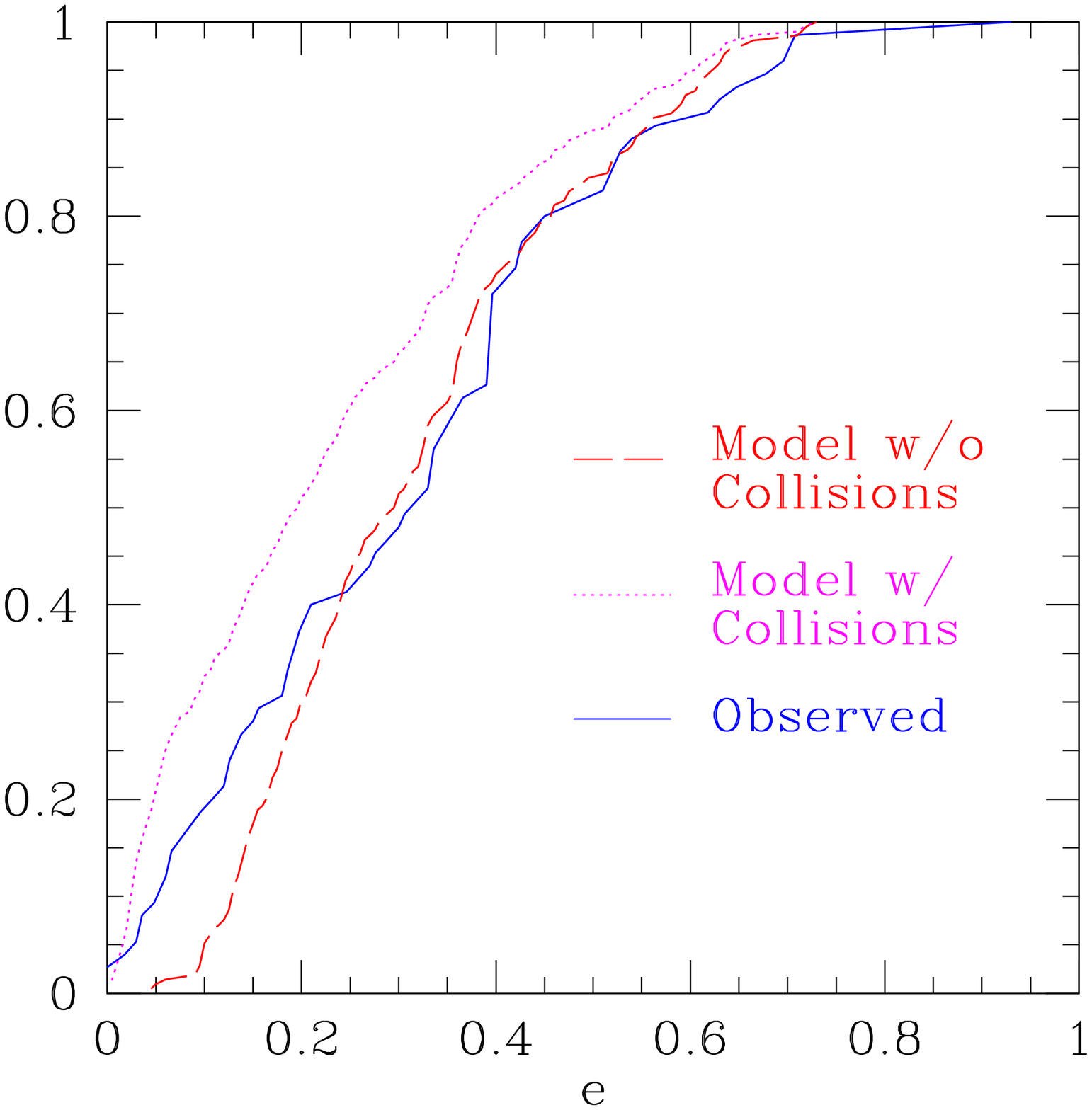}{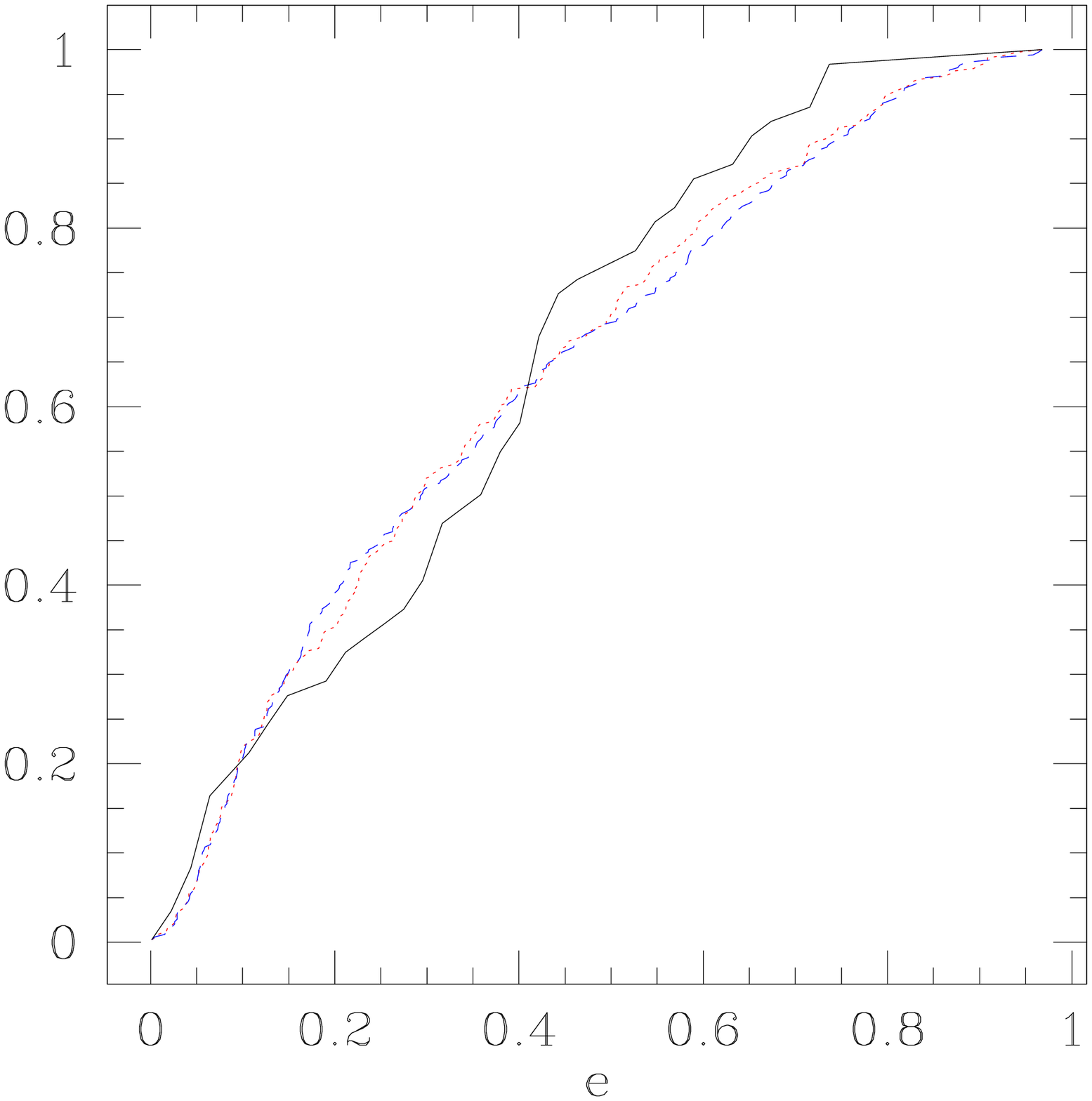}
\caption{Left: Simulation with mass distribution. Here we compare the
observed distribution of eccentricities (excluding planets with $P <
10\,$d, solid line) to the eccentricity distributions produced by our
simulations when each planet's mass is drawn independently from $P(m)
\sim m^{-1}$ for $1\, M_{\rm Jup} < m < 10\, M_{\rm Jup}$.  The dashed line shows the
eccentricity of the planet remaining after an ejection, while the
dotted line includes systems resulting in a collision.
Right: Results from simulations for 3 equal-mass planets. Here we compare the
observed distribution of eccentricities (excluding those with $P < 10\,
$d, solid black line) to the eccentricity distribution of the
remaining inner planet after there has been one ejection or collision
for one of two sets of simulations with different initial
eccentricities.  The dashed blue and dotted red lines are for simulations
with $e_{\rm init} = 0.$ and $e_{\rm init} = 0.05$, respectively.}
\end{figure}

\section{Three Planets, Equal Masses}

We have begun performing simulations of unstable systems containing three
equal-mass planets using the conservative Bulirsch-Stoer integrator in
Mercury (Chambers 1999).  The results presented here are based on
$\sim 10^3$ numerical integrations.
For three-planet systems, we set the radius of each planet at the
beginning of the integration.  For determining when collisions
occurred, each planet was assigned a radius of $R_i = R_{\mathrm Jup}$.
A planet was removed from the system if its separation from the
central star became greater than $1000$AU (ejected) or less than
$r_{\rm min} = R_\odot$ (collided with star).

For three-planet systems, a single ejection or collision results in a
system containing two planets.  The remaining two planets may either
be dynamically stable or undergo a subsequent ejection or collision
to end with a single planet.  Each run was terminated when two of the
planets were removed by being ejected, colliding with another planet,
or colliding with the star, or when the maximum integration time was reached
$t_{\rm max}= 2 \times 10^6$ years.

Our numerical integrations were performed for a system containing
three planets, each with a mass ratio $m/M \simeq 10^{-3}$.  The initial
semi-major axes are fixed at $1$, $1.39$, and $1.92$ AU, so that the
ratio of orbital periods is $P_3/P_2 = P_2/P_1 = 1.63$.  Each planet
is initially placed on a nearly circular orbit with an eccentricity of
either $0.0$ or $0.05$ and an inclination of $\pm 1^\circ$.  All
remaining angles (longitudes and phases) were randomly chosen between
$0$ and $2\pi$.  

The typical time for instability to develop is $\sim10^6$ years.  In
the majority of cases, two planets were left in a stable configuration
after one planet was ejected or two planets collided.  In Fig.\ 3
(right) we show the distributions of the eccentricity of the inner
planet after a collision or ejection of one planet.  It is in
reasonable agreement with the observed distribution for extrasolar
planets.  Both the eccentricity distribution (Fig.\ 3, right) and the
corresponding distribution of the inclination of the inner planet
(Fig.\ 4, left) are in qualitative agreement with those presented by
Marzari \& Weidenschilling (2002, based on slightly different initial
conditions).  In Fig.\ 4 (right) we show the cumulative distribution
for $\Delta \varpi = \left|\left(\Omega_2+\omega_2\right) -
\left(\Omega_1+\omega_1\right)\right|$, the angle between the
periapses of the two planets remaining after one planet has been
ejected from the system.  We do not find the two remaining planets to
have preferentially aligned periapses (cf.\ Malhotra 2002).

\begin{figure}[htb]
\plottwo{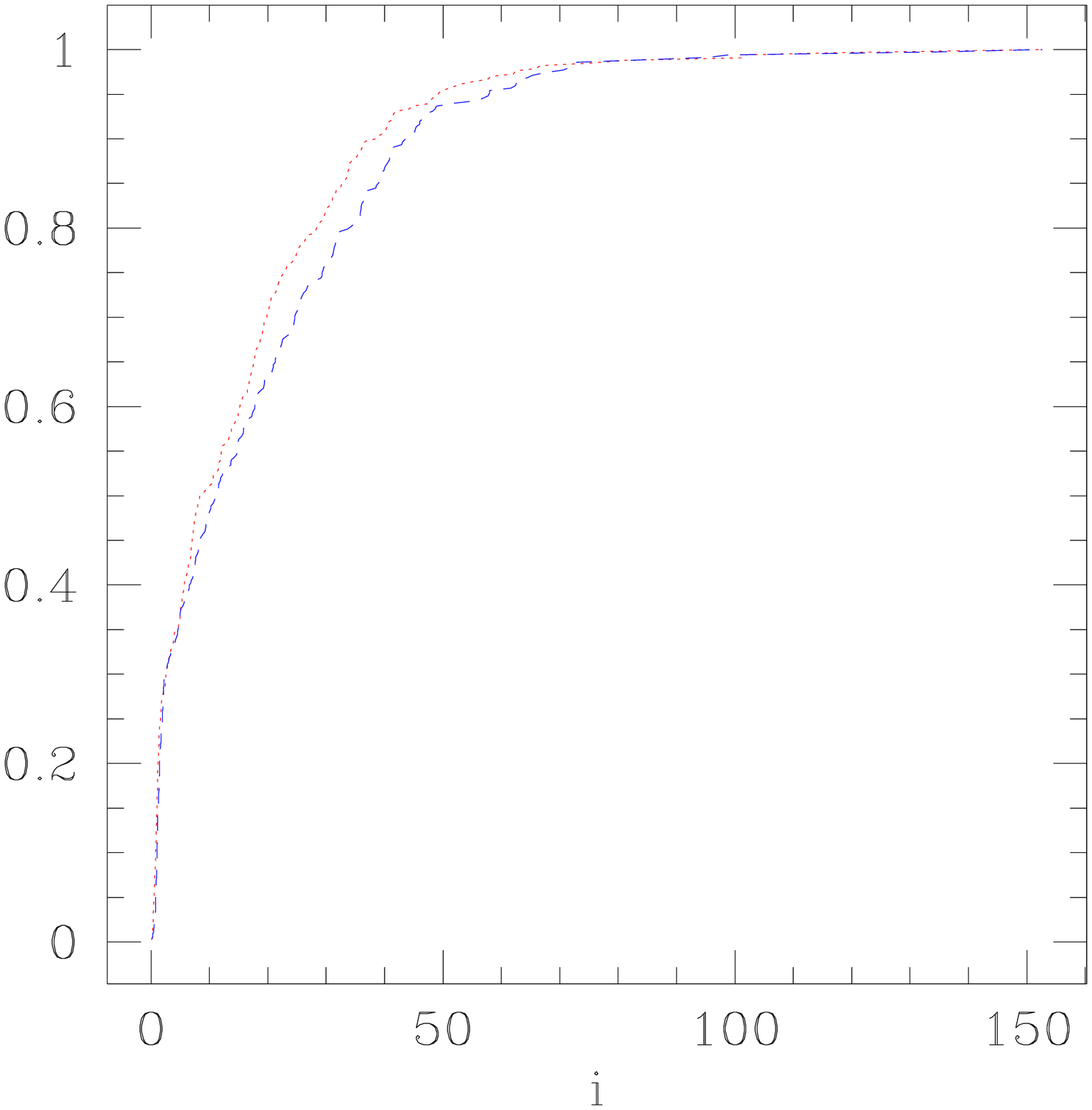}{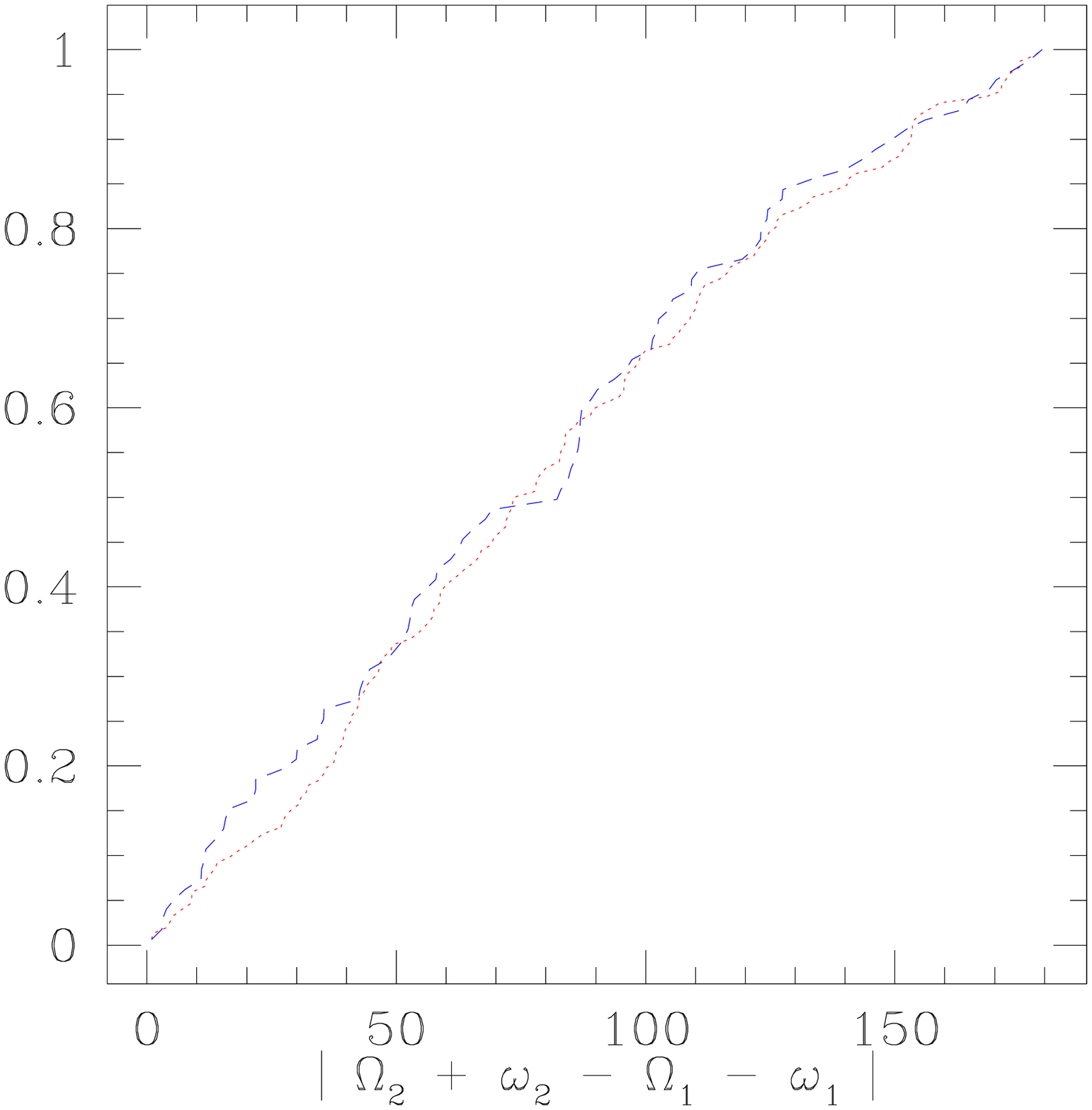}
\caption{Left: Cumulative distribution of inclination. Each line shows
the distribution of the inclination of the remaining inner planet
(with respect to the initial orbital plane) after there has been one
ejection or collision for one of two sets of simulations with
different initial eccentricities.  The dashed blue and dotted red lines are for simulations
with $e_{\rm init} = 0.$ and $e_{\rm init} = 0.05$, respectively..
Right: Cumulative distribution of the periapse misalignment angle
between two retained planets. }
\end{figure}

\section{Conclusions}

Dynamical interactions between planets of unequal mass reduces the
frequency of collisions as compared to scattering between equal-mass
planets.  In particular, planet-planet scattering can reproduce the
observed distribution of eccentric orbits using a plausible
distribution of mass ratios.  To determine if collisions are
sufficiently rare will require a more careful comparison to the
observations.  Independent of the distribution of planet masses,
planet-planet scattering by two planets initially on nearly circular
orbits produces a maximum eccentricity of $\sim 0.8$.  While this compares
favorably with the presently known extrasolar planets, future
detections will test this prediction of the two-planet scattering
model. Alternatively, the observed properties of extrasolar planets
could also be produced as a result of dynamical instabilities
developing in three-planet systems.

\acknowledgments

We are grateful to Eugene Chiang, Renu Malhotra, and Scott Tremaine for
valuable discussions.
E.B.F.\ acknowledges the support of the NSF
graduate research fellowship program and Princeton University, and thanks
the Theoretical Astrophysics group at Northwestern University for
hospitality.  F.A.R.\
and K.Y.\ acknowledge support from NSF grant AST-0206182.
Our computations were supported by the National Computational Science
Alliance under Grant AST980014N and utilized the SGI/Cray Origin2000
supercomputers at Boston University.

\end{document}